\documentstyle[aps,epsf]{revtex}  

\begin{document}        

\baselineskip 14pt
\title{Current Status of VHE Astronomy}
\author{Gus Sinnis}
\address{Los Alamos National Laboratory}
%
\maketitle              

\begin{abstract}        
Very-high-energy astronomy studies the Universe at energies between 30 GeV 
and 100 TeV.  The past decade has seen enormous progress in this field.
There are now at least seven known sources of VHE photons.  
By studying these objects in the VHE regime one can begin to understand the 
environments surrounding these objects, and how particle acceleration is 
realized in nature.  In addition the photon beams from the extragalactic 
gamma-ray sources can be used to study the electromagnetic fields in the 
intervening space.  This recent progress can be traced to the development 
of a new class of detector with the ability to differentiate between air 
showers produced by gamma rays and those produced by the much more 
numerous hadronic cosmic-ray background.  Much more sensitive instruments 
are currently in the design phase and two new types of instruments are
beginning to take data.  In this paper we will discuss the physics of these
sources and describe the existing and planned detectors.  

\end{abstract}   	

\section{Introduction}               
The first telescope sensitive to TeV radiation was an air Cerenkov telescope 
built in the 1950's by J.V. Jelley \cite{Jelly}.  It would take over 
three decades to convincingly detect a cosmic source of TeV 
photons \cite{lamb_90}.  Real progress did not occur until a second generation
of telescopes was built that could differentiate between air showers produced
by gamma rays and those produced by protons and heavier nuclei.  
We now know of at least seven sources of TeV gamma rays, from the Crab
nebula, to the distant active galaxies.  
There are several experiments now in the design phase
that promise a sensitivity more than an order of magnitude better than
current instruments.  In addition there are two entirely new classes
of detectors being brought online.  One of these is 
sensitive to very low energy gamma rays (30-50 GeV) and the other capable 
of continuously monitoring the entire overhead sky.  The next decade will 
be an exciting one for VHE astronomy.

\section{The Detection of VHE Gamma Rays}
When a high energy cosmic ray (photon or nucleus) enters the 
atmosphere it loses energy through
interactions with the nuclei in the atmosphere.  At high energies the energy
loss is dominated by processes that create particles (nuclear interactions,
bremsstrahlung, and pair creation).  The result is an extensive air shower - 
a swarm of particles traveling towards the ground at nearly the speed of 
light.  The number of particles in the swarm increases until the 
average energy per particle falls below $\sim$80 MeV.  At that point the 
dominant interactions do not lead to particle production, and the air 
shower begins to die.  For sufficiently energetic cosmic rays enough 
particles reach the ground to be directly detected.  
	
There are two well developed techniques for detecting $\sim$TeV gamma rays.
The first detects the Cherenkov radiation from the electromagnetic particles
as they traverse the atmosphere.  This radiation is beamed forward 
with an opening angle of $\sim$1 degree.  When the Cherenkov light reaches sea 
level it is in the shape of a pancake, roughly 200 meters across and 1 meter
thick.  The photon density at sea level is roughly 
200 photons m$^{-2}$ TeV$^{-1}$ of primary gamma-ray energy.  This light must 
be detected over the night-sky background, 
$\sim$4000 photons m$^{-2}$ ns$^{-1}$ sr$^{-1}$.  Air 
Cherenkov telescopes consist of one or more large mirrors which focus the
Cherenkov light onto a camera placed at the focal plane.  Present
cameras consist of an array of photomultiplier tubes (PMTs).  Since these are
pointed optical instruments, they view only a small piece of the sky 
($\sim 4\times 10^{-3}$sr), and they can only operate during clear moonless 
nights.  (Recently progress has been made in the ability to operate with 
a partial moon \cite{hegra_1}.)  Table \ref{table_detectors} 
gives the important characteristics of some of the air Cherenkov
instruments now in operation.

The second technique is to directly detect the particles that survive to 
the ground.  Until recently the dominant detector material was plastic 
scintillator.  Recently, water has been used as a detection medium.  
Since the particles also travel in a pancake
and are beamed forward, the plane of the particle front is perpendicular to the
direction of the primary cosmic ray.  Thus, by measuring the relative arrival
time of the shower front on the ground, one can reconstruct the direction
of the primary cosmic ray.  Typical scintillation arrays deployed between 
100 m$^{2}$ and 1000 m$^{2}$ of scintillator dispersed over $10^{4}$ m$^{2}$ 
to $10^{5}$ m$^{2}$ of land. 
With such a sparse sampling of the surviving particles they are only sensitive 
to much higher energy primary gamma rays ($\sim$100 TeV).  
Recent advances, such as the use of water as a detection medium 
(which allows for a dense sampling
of the air shower) and the move to very high altitudes has led to energy
thresholds near or below 1 TeV for particle detection arrays.  These particle
detector arrays view the entire overhead sky and can operate continuously.

The current set of available instruments may be divided into three categories
based on  both the experimental techniques employed and the physics goals of
the instruments: high sensitivity instruments, low-energy threshold 
instruments, and open aperture/high duty cycle instruments. 

\subsection{High Sensitivity Instruments}
The Whipple gamma-ray telescope 
was the first instrument in this class.
The mirror area is of moderate size and the camera is an array
of 330 PMTs.  Each PMT views 0.25$^{\circ}$ of the sky and the 
field-of-view of the camera is roughly 3$^{\circ}$.  More modern
instruments have been built with finer resolution cameras.  The CAT telescope 
in the Pyrenees has a camera with 0.1$^{\circ}$ pixel size.  The finer
camera resolution leads to better rejection of the background, however simulations
indicate that resolution finer than 0.1$^{\circ}$ will not lead to an improvement
in sensitivity.

This type of instrument is known as a high-resolution imaging detector.
The shape and orientation of the image of the Cherenkov 
light pool is sensitive to the 
direction and type of the primary cosmic ray.  The image induced by a primary 
gamma ray forms an ellipse.  The semimajor axis of the ellipse
points to the source position in the image plane.  In contrast,
the image of a proton 
initiated shower will be more chaotic and the larger transverse
momentum of the hadronic interactions leads to a wider image.  Also, in general
the protons do not come from the source direction, hence the semimajor axis of
a fit ellipse will not point to the source location.  

The energy resolution of these instruments is limited by 
their ability to locate the "core" of the extensive air shower.  
(The core of an EAS is the location where the primary cosmic ray 
would have hit the ground in the absence of an atmosphere.)  
At present the HEGRA collaboration has an array of six telescopes 
operating in the Canary Islands.  Multiple measurements  the shower density 
allow one to reconstruct the core location with high precision.  The energy 
resolution of this array is roughly 20\% ($\Delta$E/E), while
the energy resolution of single telescope is $\sim$40\%.

\subsection{Low Energy Threshold Instruments}
Huge mirror areas must be used to achieve energy thresholds well below 100 GeV.  
Since the Cherenkov light must be detected in the presence of the night-sky 
background, the energy threshold of an air-Cherenkov telescope is inversely 
proportional to the square root of the mirror area.  In the early eighties 
the large mirror area afforded by solar power plants was suggested as an 
economical way of achieving very low energy thresholds (\cite{tumay_1}).  
At present there are two such installations beginning operations,  
the CELESTE group operating in the Pyrenees and the STACEE collaboration 
in the United States.  Both groups have achieved energy thresholds below 50 GeV.

\subsection{Large Aperture/High Duty Factor Instruments}   

Despite the relatively high energy thresholds of these instruments they have several 
advantages over air-Cherenkov telescopes.  They can be operated 
continuously and they view the entire 
overhead sky.  There have been two efforts to lower the 
energy threshold of these instruments.  The Tibet collaboration 
has built a small, dense scintillator array at very high altitude (4 km asl).  At this 
altitude many more shower particles survive to the ground.  They have achieved an
energy threshold below 10 TeV.  The second group has developed a water Cherenkov 
detector, known as Milagro \cite{sinnis}.  Since the Cherenkov angle in water is 
$\sim41^{\circ}$, 
a sparse array of PMTs in a large pool of water can detect nearly every electromagnetic 
particle that reaches the pool.  In addition the gamma rays in the air shower (which 
outnumber the electrons by $\sim$4:1) convert to electrons and/or positrons in the
water and are detected with high efficiency.  Using the dense sampling of the air shower 
the Milagro collaboration hopes to have an energy 
threshold below 1 TeV.

\section{The Physics of VHE Gamma Rays}

\subsection{Neutron Stars}
Neutron stars are compact objects, measuring $\sim$10 km across with a mass of about 
1.4 times the mass of our sun (M$_{\odot}$).  The surface layer is composed of 
iron and the density near the core is $\sim$10$^{15}$ gm cm$^{-3}$.  They are 
the final stages of stellar evolution, being left behind after a massive star goes 
supernova.  Conservation of angular momentum and magnetic flux lead 
to rotational periods of several milliseconds to 
10s of seconds and magnetic fields of order 10$^{12}$ Gauss at the surface.  

The region around the neutron star is known as the magnetosphere, which
is composed of a charge separated plasma threaded by magnetic field lines.  
Charged particles escape the magnetosphere along the open field lines 
emanating from the poles of the star.  This wind of particles terminates in a shock
where electrons are accelerated to TeV energies.  The energetic electrons enter 
the nebula, emit synchrotron radiation as they spiral in the magnetic field, 
and upscatter the synchrotron photons to TeV energies via inverse Compton 
scattering.  This model is known as the synchrotron self-Compton (SSC) 
model \cite{harding_1}.

\subsection{Active Galaxies}

The heart of an active galaxy is a supermassive 
black hole ($10^{6}-10^{10}$ M$_{\odot}$).  The black hole is surrounded by an accretion 
disk that is fed by the host galaxy.  Radio loud AGN 
have jets of relativistic particles emitted along their rotation axes.  Among the radio 
loud AGN it is believed that all the observed classes are simply the same type of object 
viewed from different orientations.  Blazars are a special class of radio loud AGN
that have their jets closely aligned with the line-of-sight.  To date all of the AGN
detected in the VHE regime are blazars.

There are two general models for the production of TeV gamma rays from AGN.  
In the first model, similar to the SSC model detailed above,  
electrons are accelerated by a shock front moving within the jet.  
These electrons emit synchrotron radiation which they then 
upscatter to TeV energies.  Variants to this model exist, the main 
variation being the source of the upscattered photons. 

In the second class of models 
protons are accelerated to extremely high energies, $10^{18}$ eV.  
These protons initiate a 
particle cascade through $p\gamma\rightarrow\Delta\rightarrow\pi 's$ 
(the ambient $\gamma 's$ come from synchrotron radiation emitted by
accelerated electrons).  The
$\pi^{0}$'s decay into photons which pair produce with the same
synchrotron photons.  The resulting electromagnetic cascade shifts the
gamma-ray energy to $\sim$10 TeV \cite{mannheim}.

Since electrons are light they can be accelerated quickly, 
though radiation losses make it difficult to accelerate them to very high
energies.  Conversely, protons can be accelerated to very high energies, 
but their acceleration rate is slower. 
Both models predict a characteristic two-humped energy spectrum.  
In the SSC model X-rays are produced by synchrotron emission from energetic electrons.  
TeV gamma rays are then produced via inverse Compton scattering of the X-ray
photons and the same population of electrons.  

\subsection{Gamma-Ray Bursts}
Gamma ray bursts are short intense bursts of gamma rays. 
They vary in duration from milliseconds to 
hundreds of seconds.   
While perhaps not universally accepted, the preponderance of evidence indicates that
GRBs are distant phenomena, laying well outside of our galaxy.  Most current models
involve one or more highly relativistic shells (perhaps the aftermath of a 
neutron star-neutron star merger) interacting with each other or with an external 
medium.  A highly relativistic shock forms at the interaction region, leading
to particle acceleration \cite{mezaros,piran}.  Though no detections of GRBs 
have yet been made in the TeV regime, the EGRET instrument (on board the Compton 
Gamma Ray Observatory) has observed
GeV photons from GRBs \cite{hurley}.  The Milagro and Tibet experiments 
(with their all-sky, all-time capability) will greatly increase our sensitivity 
to TeV emission from GRBs.

\subsection{The Inter-Galactic Infra-Red Radiation Field}

The interaction cross-section for $\gamma\gamma\rightarrow e^+e^-$ is large when the 
invariant mass of the two 
colliding photons is just above twice the rest mass of the electron. TeV photons 
interact with infrared and visible photons, leading to an apparent attenuation of the 
source at high energies \cite{gould}.  The 
attenuation of high-energy photons depends on the distance to the source and the 
intensity of the inter-galactic infrared radiation (IGIR) field.  Since most of 
this field has its origin in the early period of star formation, it should be 
sensitive to the details of structure formation in the Universe \cite{macminn}. 
However, significant reprocessing of this radiation has occurred over the lifetime 
of the Universe.  At present it is still a matter of debate if the initial conditions 
of structure formation can be extracted from a measurement of the IGIR 
field \cite{stecker}.  

In principle one could use the observed spectra of AGN to measure the IGIR field.  
However, there is the complication that absorption of TeV photons may also occur at the 
source, and we do not know the intrinsic energy spectrum of these sources.  
To improve the models of particle acceleration in AGN,
multi-wavelength campaigns (TeV, x-ray, and optical) of AGN in both their 
flaring and quiescent states are required.   However, the observation of similar AGN at 
several redshifts will probably be required before a 
convincing measurement of the IGIR can be made.  Since only weak AGN have been detected 
to date in the TeV regime, this implies the need for telescopes with sensitivities 
$\sim$10 times greater than current instruments.  Several such instruments are planned.

\section{Observations of VHE Gamma-Ray Sources}

\subsection{Galactic Sources}
\begin{figure}[ht]
\centerline{\epsfxsize 3.1 truein \epsfbox{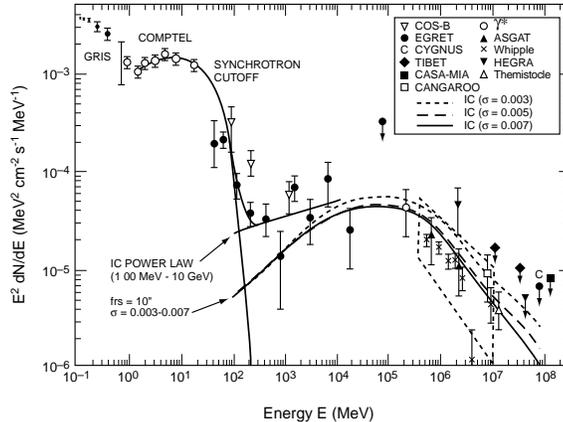}}
\caption[]{\label{fig_crabspec}
\small The measured energy spectrum of the Crab nebula.  Predictions of the SSC 
model are superimposed on the figure.  The different model predictions correspond to 
different values for the ratio of magnetic energy to the total energy ($\sigma$) 
in the pulsar wind.}
\end{figure}

The Crab nebula is powered by a pulsar spinning 30 times per second.  
The nebula is filled with
relativistic electrons injected from the pulsar.  Figure \ref{fig_crabspec} shows the 
measured energy spectrum of the Crab out to almost 50 TeV.  The high energy measurements
have been made by the CANGAROO collaboration by observing the source
at large zenith angles \cite{tanimori_1}.  The data are well described by the SSC model 
outlined above \cite{harding_1}.  The best fits to this model indicate that the magnetic
field strength in the region where the TeV gamma rays are produced is roughly 16 nT 
\cite{hillas_1}.  
The parameter $\sigma$ in the figure is the ratio of magnetic energy to total energy 
in the pulsar wind.  The bulk of the energy in the wind is the kinetic energy of 
the particles.  To date, no pulsed emission from the Crab has been observed above 30 GeV.

Three other galactic sources have been detected in the TeV regime: 
Vela \cite{yoshikoshi}, PSR1706-44 \cite{kifune}, and SN1006 \cite{tanimori}.
The first two of these sources are filled supernova remnants similar to the 
Crab nebula, while SN1006 is a shell type supernova remnant (SNR).  SNRs are 
the favored location for the acceleration of hadronic cosmic rays.  The TeV 
emission emanates from one rim of the expanding shell.  While this discovery 
at first appeared to be evidence of cosmic ray acceleration within supernova 
remnants, recent calculations indicate that the observed TeV emission is 
consistent with expectations from an inverse Compton component associated 
with the acceleration of electrons.  Thus there is still no direct evidence for the 
acceleration of nuclei within supernova remnants.

\vskip -4.5 cm

\subsection{Extra-Galactic VHE Gamma-Ray Sources}
To date two confirmed detections of AGN in the TeV energy regime: Mrk421 and Mrk501. 
There are also several unconfirmed AGN detections: 1ES 2344+514 and 
PKS 2155-304 \cite{turver}. 
Mrk421, Mrk501, and 1ES 2344+514 are relatively close with redshifts of 
order 0.03, while PKS 2155-304 lies at a redshift of 0.117.
 Mrk421 was the first extragalactic object to be observed in the TeV band.  
The average flux from this source is 
roughly 1/3 that of the Crab nebula \cite{punch_1}.  However, in contrast to the 
emission from the Crab, the luminosity of Mrk421 varies wildly over a large 
range of timescales.  This same variability has also been observed in Mrk501.
The AGN 1ES 2344+514 was discovered during an apparent flare, however the 
quiescent emission is too low to monitor the long-term behavior of this 
source \cite{catanese}.  Observationally there are two important measurements 
made of the AGN, their variability and their energy spectra.

\begin{figure}[ht]
\centerline{\epsfxsize 3.0 truein \epsfbox{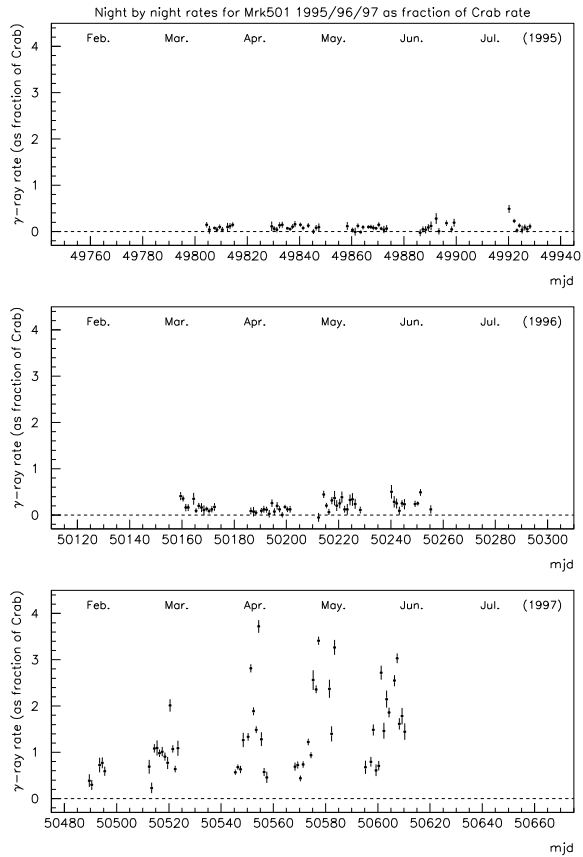}}   
\vskip -.0 cm
\caption[]{
\label{fig_mrkvar}
\small The measured flux from Mrk501 over a three year interval \cite{quinn}.}
\end{figure}

The observed minimum time scale of variability is related to the size of the source, 
the relativistic bulk motion of the source,
and the acceleration timescale. The shortest variability timescale
observed to date, from Mrk421 with the Whipple observatory \cite{gaidos_1}, was 
fifteen minutes.   A short variability timescale is
more easily explained in electron models, although the proton model can explain the 
current observations if the bulk Lorentz factor of the source is increased.  
Also of interest is the long term flaring behavior of these sources.  In Figure 
\ref{fig_mrkvar} we show the observed flux from Mrk501 over a three year period.  
The source exhibits long periods of inactivity followed by periods of intense 
flaring that may last for several years.  The cause
of this long term behavior is as yet unknown.  

The upper range of the energy spectrum of an AGN is determined by the bulk Lorentz 
factor, the balance between acceleration and cooling of the particles, and the 
optical depth of the source to TeV gamma rays.  The best measurement of the 
energy spectrum of an AGN above 5 TeV has been made by the HEGRA group \cite{hegra_2}. 
While the HEGRA group has a measured a significant number of photons between 
19 and 24 TeV, they establish a 2$\sigma$ C.L. lower limit to the maximum 
energy of 16 TeV.  Since electrons cool more efficiently than protons it is 
more difficult to accelerate them to very high energies.  However, current 
measurements of the maximum energy are consistent with the SSC model.

\section{The Future of VHE Astronomy}
TeV gamma-ray astronomy has reached a productive stage of development.  
The 1970's and early 1980's were a period marked by controversy and conflicting 
results.  The development of a second generation of imaging Cherenkov telescopes 
paved the way for convincing discoveries of sources of TeV gamma rays.  
Several new instruments just now coming online promise to open the field 
to new types of physics.  All-sky instruments such as Milagro and Tibet 
will be sensitive to a TeV component of gamma-ray bursts.  
They also have the ability to provide long-term monitoring of AGN and perhaps discover 
new sources of TeV gamma rays.  The solar array air Cherenkov telescopes will  
have very low energy thresholds, comparable to the upper energy range of 
space-based instruments.  They could enable us to observe a pulsed component 
to the Crab, observe the more distant AGN, and/or AGN where absorption of TeV 
photons at the source is important.  

\begin{figure}[ht]
\centerline{\epsfxsize 2. truein \epsfbox{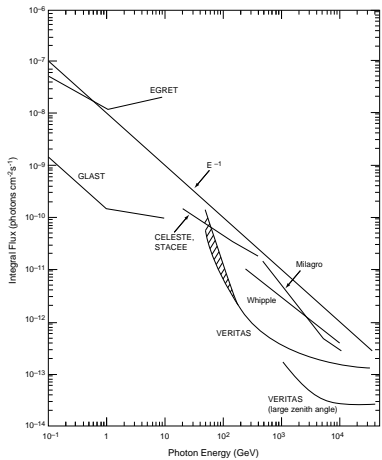}}   
\vskip -.0 cm
\caption[]{
\label{fig_sensitivity}
\small The sensitivity of selected current and planned TeV gamma-ray observatories.}
\end{figure}

Farther down the road there are several
instruments planned that will achieve sensitivities over ten times better than current 
instruments.  These include the Veritas array and MAGIC in the northern hemisphere and
HESS and Super CANGAROO in the southern hemisphere.  
Veritas \cite{veritas}, HESS, and Super CANGAROO will consist of 
arrays of large high-resolution imaging telescopes.  The MAGIC telescope will be one
large 220 m$^2$ mirror with high quantum efficiency PMTs in the camera. 
Figure \ref{fig_sensitivity} shows the expected sensitivity of several current and 
planned instruments.  These instruments should detect over 30 AGN and allow for a full
exploration of the TeV Universe.  

\vskip -2.5 cm

\vskip -0.5 cm
\begin{table}
\caption{\label{table_detectors}
Characteristics of current VHE air Cherenkov telescopes.  
Numbers given in parentheses are planned.}
\begin{tabular}{lccccc} 
 Detector  &  Mirror Area & Energy Threshold (GeV) & Number of Pixels & Number of Telescopes & Location\\  
 \tableline
Whipple  & 75 m$^2$           &  200   &  330  & 1 & Arizona, USA \\ 
\tableline
CAT      & 17 m$^2$           &  190   &  534  & 1 & Pyrenees, France \\
\tableline
CANGAROO & 11.3 m$^2$         &  1000  &  256  & 1 & Woomera, Australia \\
\tableline
HEGRA    & 8.5 m$^2 \times$ 5 &  1000  &  271$\times$5 & 5 & La Palma, Canary Islands \\
\tableline
Telescope Array & 6 m$^2 \times$ 7 & 600    & 256     &  7 & Dugway, Utah \\
\tableline
Durham   & 42 m$^2 \times$ 3       & 200    & 109     &  3 & Woomera, Australia \\
\tableline
CELESTE  & 1000 (2000) m$^2$       & 60(30) & 18(40)  &  18(40) & Pyrenees, France \\
\tableline 
STACEE   & 1230 (2500) m$^2$       & 75(50) & 32(64)  & 32(64)  & New Mexico, USA \\
\end{tabular}
\end{table}
\nobreak

\end{document}